\documentclass[%
reprint,
superscriptaddress,
amsmath,amssymb,
aps,
prapplied,
]{revtex4-1}  

\usepackage{graphicx}
\usepackage{dcolumn}
\usepackage{bm}

\begin{document}
	
	\preprint{APS/123-QED}

\title{Designing large arrays of interacting spin-torque nano-oscillators for microwave information processing}

\author{P. Talatchian}
\affiliation{Unit\'e Mixte de Physique, CNRS, Thales, Univ. Paris-Sud, Universit\'e Paris-Saclay, 91767 Palaiseau, France}
\author{M. Romera}
\affiliation{Unit\'e Mixte de Physique, CNRS, Thales, Univ. Paris-Sud, Universit\'e Paris-Saclay, 91767 Palaiseau, France}
\author{F. Abreu Araujo}
\affiliation{Institute of Condensed Matter and Nanosciences, UCLouvain, Place Croix du Sud 1, 1348 Louvain-la-Neuve, Belgium}
\author{P. Bortolotti}
\affiliation{Unit\'e Mixte de Physique, CNRS, Thales, Univ. Paris-Sud, Universit\'e Paris-Saclay, 91767 Palaiseau, France}
\author{V. Cros}
\affiliation{Unit\'e Mixte de Physique, CNRS, Thales, Univ. Paris-Sud, Universit\'e Paris-Saclay, 91767 Palaiseau, France}
\author{D. Vodenicarevic}
\affiliation{Centre de Nanosciences et de Nanotechnologies, CNRS, Univ. Paris-Sud, Universit\'e Paris-Saclay, 91405 Orsay, France}
\author{N. Locatelli}
\affiliation{Centre de Nanosciences et de Nanotechnologies, CNRS, Univ. Paris-Sud, Universit\'e Paris-Saclay, 91405 Orsay, France}
\author{D. Querlioz}
\affiliation{Centre de Nanosciences et de Nanotechnologies, CNRS, Univ. Paris-Sud, Universit\'e Paris-Saclay, 91405 Orsay, France}
\author{J. Grollier}
\affiliation{Unit\'e Mixte de Physique, CNRS, Thales, Univ. Paris-Sud, Universit\'e Paris-Saclay, 91767 Palaiseau, France}


\begin{abstract}
	Arrays of spin-torque nano-oscillators are promising for broadband microwave signal detection and processing, as well as for neuromorphic computing. In many of these applications, the oscillators should be engineered to have equally-spaced frequencies and equal sensitivity to microwave inputs. Here we design spin-torque nano-oscillator arrays with these rules and estimate their optimum size for a given sensitivity, as well as the frequency range that they cover. For this purpose, we explore analytically and numerically conditions to obtain vortex spin-torque nano-oscillators with equally-spaced gyrotropic oscillation frequencies and having all similar synchronization bandwidths to input microwave signals. We show that arrays of hundreds of  oscillators covering ranges of several hundred MHz can be built taking into account nanofabrication constraints.

%
\end{abstract}

\maketitle

\section{Introduction}

%

Spin-torque nano-oscillators 
\cite{kiselev_microwave_2003,rippard_direct-current_2004} are nanoscale magnetic tunnel junctions composed of two ferromagnetic layers separated by a thin non-magnetic layer (Fig. 1a). They have the same structure as the actual cells in magnetic non volatile memories, and can be fabricated in large numbers in microelectronic chips \cite{chung_4gbit_2016}. A current applied to these nanojunctions becomes spin 
polarized and applies a spin-transfer torque on the local magnetization \cite{slonczewski,berger}. For current densities above a threshold, this torque can generate sustained oscillations of the free layer magnetization, which in turn are converted into microwave voltage oscillations through magnetoresistive effects. The frequency of the oscillations can be varied from hundreds of MHz to tens of GHz by changing the materials and the geometry of the junctions in order to select the modes of magnetization dynamics \cite{bonetti_spin_2009}. Interestingly with spin-torque oscillators, once the pillar is fabricated, the frequency can still be tuned by hundreds of percent by varying the applied direct current or the magnetic field \cite{slavin_nonlinear_2009}. Spin-torque nano-oscillators  also respond to input microwave signals in a large frequency band around frequencies at which they oscillate. This  response can take multiple forms. For example, spin-torque nano-oscillators generate direct voltages if the input is a microwave current with a frequency close to their own. This rectification is called spin-diode effect \cite{miwa_highly_2014,jenkins_spin-torque_2016}. Another response for spin-torque nano-oscillators in the auto-oscillation regime is the synchronization of their oscillations to input microwave signals on a frequency span called injection locking range that can reach several percent of their base frequency \cite{rippard_injection_2005,LebrunPRL}. 

 We can therefore envision using arrays of spin-torque nano-oscillators with different base frequencies to analyze or process microwave signals on wide frequency bands from MHz to GHz \cite{ebels_spintronic_2017,louis_low_2017} (Fig. 1b). The advantages of these circuits compared to standard spectral analysis techniques are the speed of processing, naturally performed in parallel, and the small dimensions of the arrays, based on nanoscale components with native sensitivity to microwaves and demonstrated CMOS compatibility. For example, in prior work, a small hardware array of four nano-oscillators has been built and used as a neural network for classifying microwave inputs in a range of a few tens of MHz \cite{romera_vowel_2017}. In order to scale these experiments to practical applications, this frequency band needs to be adapted. For some applications, depending on the input to analyze, this frequency band will need to be increased to hundreds of MHz or more. For other applications it will be on the contrary more important to increase the frequency sensitivity than the covered frequency band. In both cases, this can be achieved by increasing the number of oscillators in the array and carefully choosing their properties.
 
 \begin{figure}
 	\includegraphics[scale=0.25]{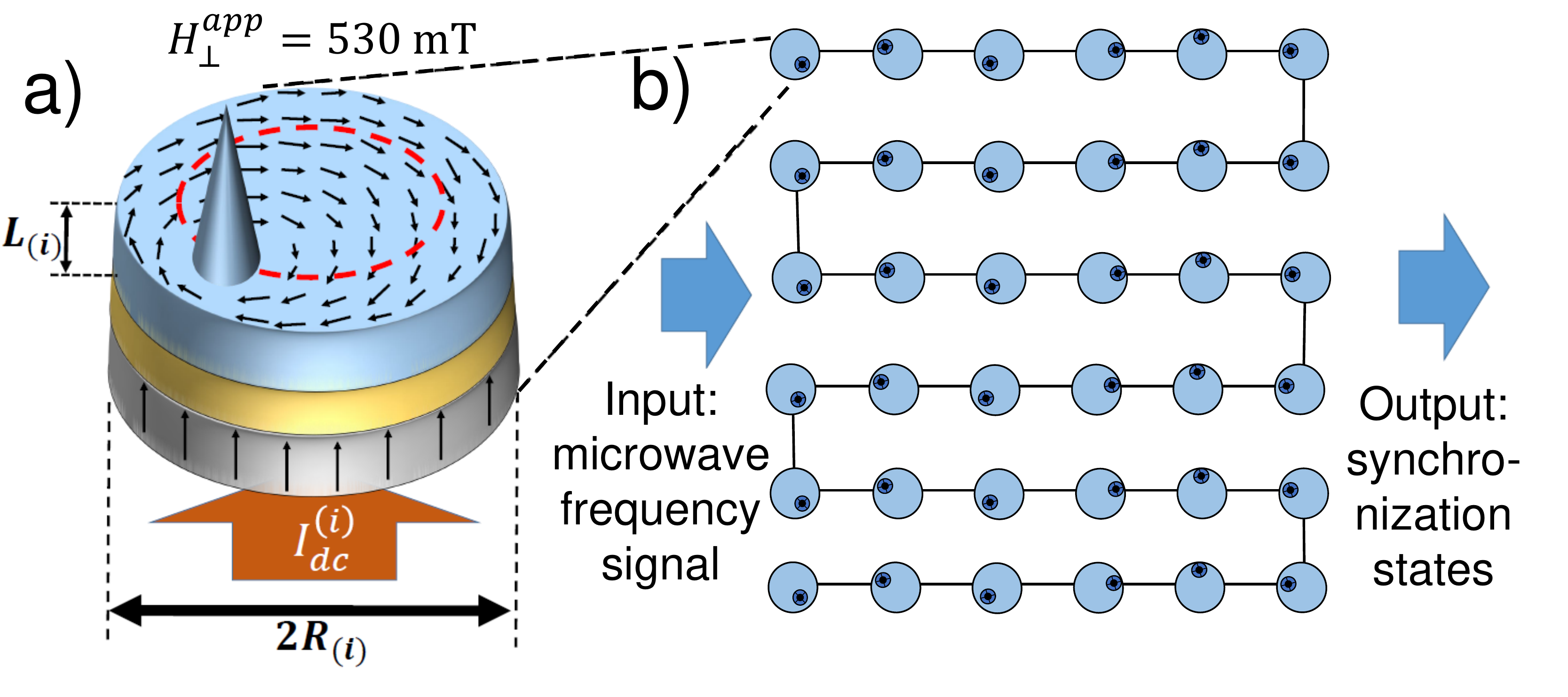}
 	\caption{\label{fig:epsart1} a) Schematic illustration of the spin-torque nano-oscillator having a magnetic vortex configuration for the free layer (blue). The yellow layer illustrates the non-magnetic layer and the gray layer corresponds to the pinned layer. The magnetization of the free layer is planar except in the vortex core area where it becomes out of plane. b) The schematic illustrates an array of N interacting spin-torque nano-oscillators receiving microwave frequency inputs (for more details see Appendix \ref{app:ElectricalCircuit}). The different synchronization states of this array correspond to the output.  }
 \end{figure}

 The fabrication of such arrays is a major challenge towards many envisaged applications based on spin-torque nano-oscillators, but their design has never been investigated. It requires finely tuning the base frequency of the oscillators and the bandwidth of their response which both depend in different ways on the same parameters: injected direct current and geometry of the pillars. Furthermore, it is important to check that the mutual coupling between oscillators, which naturally arises when they are electrically connected or closely packed, does not compromise their individual response \cite{awad_long-range_2017,Lebrun_Mutual,locatelli_efficient_2015}. In this work, our focus is on oscillators with a vortex in the free layer as their properties are well described and understood \cite{guslienko_magnetic_2006,bortolotti_temperature_2012,grimaldi2014response}. We analytically derive design rules to build large arrays of uncoupled vortex spin-torque nano-oscillators with equally spaced frequencies and equal frequency sensitivity that can process microwave inputs on a wide frequency range. Here, by frequency sensitivity we refer to the frequency precision with which two microwave inputs can be distinguished by the array. Input microwave signals can be introduced either as ac magnetic fields (\cite{romera_vowel_2017} and this work) or as ac currents injected directly in the electrical circuit of the array \cite{romera_enhancing_2016}. We computed the optimal operating points (applied dc currents) and physical properties (size and aspect ratio) of the oscillators in the array. We show that arrays comprising hundreds of  vortex oscillators with an overall response covering hundreds of MHz can be produced with existing nanofabrication techniques. We find that, counter-intuitively, arrays with the smaller number of oscillators will have the larger overall frequency band, but at the expense of a reduced frequency sensitivity. Finally we numerically simulate an array designed with these rules, taking into account the mutual coupling between oscillators. We show that for experimentally observed coupling values the whole array is functional and that the design rules derived analytically in the absence of coupling can be applied.
 
 \section{Analytical expressions of the oscillator frequency and injection locking bandwidth}

 In order to design the arrays, the analytical expressions of the frequency and injection locking band of spin-torque nano-oscillators are first derived.  As illustrated in Fig. 1a, spin-torque nano-oscillators with a vortex configuration in the free 
 layer \cite{metlov_stability_2002} have a planar magnetization except in the vortex core area where it becomes out of plane. Here, we focus on spin-torque oscillators having a magnetic tunnel junction structure, however our approach can be extended to other magnetic structures. When a sufficient electrical current density is injected in the nano-pillar, 
 the vortex core leaves its initial position in the free layer and starts to oscillate in a quasi-circular trajectory. By solving the Thiele 
 equation \cite{thiele_steady-state_1973,dussaux_field_2012} describing the 
 trajectory of the vortex core in the steady-state, the expression of the 
 frequency of the vortex oscillations $ f^{(i)} $ can be determined in
 Eq.~(1) :
 \begin{equation}
 f^{(i)}=\dfrac{1}{2\pi 
 G^{(i)}}\{\kappa_{ms}^{(i)}+\kappa_{Oe}^{(i)}J^{(i)}+(\kappa_{ms}^{'(i)}+\kappa_{Oe}^{'(i)}J^{(i)})p_0^{(i)}\},
 \end{equation}
 with $G^{(i)}$ the gyrovector magnitude, $\kappa_{ms}^{(i)}$ the magnetostatic confinement, $\kappa_{Oe}^{(i)}$ Oersted field confinement, $\kappa_{ms}^{'(i)}$ the nonlinear magnetostatic confinement, $\kappa_{Oe}^{'(i)}$ the nonlinear Oersted field confinement (see Appendix \ref{app:OerstedMagnetoStat}) and $J^{(i)}$ the applied current density ($J^{(i)}=I_{dc}^{(i)}/\pi {\cal R}_{(i)}^2$, where ${\cal R}_{(i)}$ is the junction radius and $I_{dc}^{(i)}$ is the applied dc current) \cite{dussaux_field_2012}. This frequency depends on the power of oscillations $p_0^{(i)}$ described by Eq.~(2):
  \begin{equation} 
  p_0^{(i)}=\dfrac{\dfrac{a_j G^{(i)}}{D^{(i)}}J^{(i)}-(\kappa_{ms}^{(i)}+\kappa_{Oe}^{(i)}J^{(i)})}{\kappa_{ms}^{'(i)}+\kappa_{Oe}^{'(i)}J^{(i)}+\xi(\kappa_{ms}^{(i)}+\kappa_{Oe}^{(i)}J^{(i)})},
  \end{equation}
  with $a_j$ the spin-transfer torque efficiency, $D^{(i)}$ the damping and $\xi$ its nonlinear factor \cite{dussaux_field_2012}. The nonlinearity of the auto-oscillator is characterized by the nonlinear frequency shift $\nu^{(i)}$ \cite{slavin_nonlinear_2009,grimaldi2014response} defined by Eq.~(3):
  \begin{equation}
  \nu^{(i)}=\dfrac{G^{(i)}}{D^{(i)}}\dfrac{\kappa_{ms}^{'(i)}+\kappa_{Oe}^{'(i)}J^{(i)}}{\kappa_{ms}^{'(i)}+\kappa_{Oe}^{'(i)}J^{(i)}+\xi(\kappa_{ms}^{(i)}+\kappa_{Oe}^{(i)}J^{(i)})}.
  \end{equation} 
  This parameter  combined with the power  $p_0^{(i)}$ affects the frequency 
  injection locking range $\Delta^{(i)}$ on which the oscillator synchronizes to an external microwave signal of amplitude $F_e$. The 
  expression of the injection locking-range is given by 
  Eq.~(4) \cite{slavin_nonlinear_2009}.
  \begin{equation} 
  \Delta^{(i)}=\dfrac{\sqrt{1+{\nu^{(i)}}^{2}}}{\sqrt{p_0^{(i)}}}F_e.
  \end{equation}
  The coefficients of these equations are
  described in Tables I and II. Their values depend on the magnetic material used as a free-layer. Here we chose to use parameters for free-layers made of FeB \cite{tsunegi} (Table I).  Importantly, as can be seen in Table II, coefficients for the 
  electrical current density $J^{(i)}$, the damping $D^{(i)}$, the confinement 
  due to the Oersted field $\kappa_{Oe}^{(i)}$, the magnetostatic confinement 
  $\kappa_{ms}^{(i)}$ and the gyroforce $G^{(i)}$, depend on the free-layer 
  radius ${\cal R}_{(i)}$, thickness  $L_{(i)}$, and applied dc current $I_{dc}^{(i)}$.

\begin{table}
	\caption{\label{tab:table1}Constant parameters of the study for FeB free-layer. Here $\mu_0=4\pi\times10^{-7}{T.m.A^{-1}}$, $\hbar=1.054\times10^{-34}J.s^{-1}$, $\gamma_G=1.76\times10^{11}rad.s^{-1}.T^{-1}$ and $e=1.602\times10^{-19}C$  }
	\begin{ruledtabular}
		\begin{tabular}{l}
			$H_{\perp}=530$ mT (fixed perpendicular applied magnetic field)\\
			$M_s=1.5\times10^{6} A.m^{-1}$ (free layer  magnetization)\footnote{ For free-layer thicknesses larger than 3 nm, $M_s$ variations smaller than 5\% can be assumed. Therefore, for free-layer thicknesses used in this work (3.0 to 8.1 nm), for simplicity, a constant $M_s$ value was used.}\\
			$\alpha_G=0.0054$ (Gilbert damping)\cite{tsunegi2014damping}\\
			$A=20\times10^{-11} J.m^{-1}$ (exchange constant)\\
			$P=0.26$ (spin polarization)\\
			$M_s^{pol}=1.2\times10^{6} A.m^{-1}$ (polarizer magnetization)\\
			$\xi=0.6$ (nonlinear damping coefficient)\cite{grimaldi2014response,khvalkovskiy_nonuniformity_2010}\\
		    $\theta_0=cos^{-1}\dfrac{H_{\perp}}{\mu_0M_s}$ (free layer magnetization angle)\\
			$b=2L_{ex}=2\sqrt{\dfrac{2A}{\mu_0M_s^{2}}}$ (vortex core radius)\\
			$a_j=\pi\dfrac{\hbar P}{2e}\dfrac{H_{\perp}}{\mu_0M_s^{pol}}sin^{2}\theta_0$ (spin-transfer torque efficiency)\\
		\end{tabular}
	\end{ruledtabular}
\end{table}

\begin{table}
	\caption{\label{tab:table3}  Parameters that depend on the applied dc current $I_{dc}^{(i)}$, the free-layer radius ${\cal R}_{(i)}$ and the free-layer thickness $L_{(i)}$.}
	\begin{ruledtabular}
		\begin{tabular}{l}
			$D^{(i)}=\alpha_G(2\pi L_{(i)}\dfrac{M_s}{\gamma_G})(\dfrac{1}{2}\ln (\dfrac{{\cal R}_{(i)}}{2b})-\dfrac{1}{8})\sin ^{2}\theta_0$ (damping)\cite{dussaux_field_2012}\\
			$G^{(i)}=(2\pi L_{(i)}\dfrac{M_s}{\gamma_G})(1-\cos\theta_0)$ (gyrovector magnitude)\\
			$\kappa_{ms}^{(i)}=(\dfrac{10}{9})\mu_0 M_s^{2} \dfrac{L{(i)}^{2}}{{\cal R}_{(i)}}\sin ^{2}\theta_0$ (magnetostatic coefficient)\cite{guslienko_magnetic_2006,gaididei_magnetic_2010}\\
			$\kappa_{ms}^{'(i)}=0.25\kappa_{ms}$ (nonlinear magnetostatic coefficient)\cite{gaididei_magnetic_2010}\\
			$\kappa_{Oe}^{(i)}=0.85\mu_0 M_s L_{(i)}{\cal R}_{(i)}\sin \theta_0$ (Oersted field confinement)\cite{khvalkovskiy_nonuniformity_2010}\\
			$\kappa_{Oe}^{'(i)}=-0.5\kappa_{Oe}$ (nonlinear Oersted field confinement)\cite{khvalkovskiy_nonuniformity_2010}\\
		\end{tabular}
	\end{ruledtabular}
\end{table}
\section{Tuning individual oscillator parameters for building large arrays: design rules.}

In this section, the analytical model presented in the previous section is used to design an array of spin-torque nano-oscillators that can process microwave signals. This is achieved through their synchronization to the input microwave signals that they receive. Ideally, this microwave processing should be done on a wide range of input frequencies, with uniform sensitivity to all frequencies, and without any input frequency gap intervals where the nano-oscillator array will not be able to respond. These conditions allowed reaching the highest 
performance on a pattern classification task in experiments and in simulations for a small neural network of four spin-torque nano-oscillators \cite{romera_vowel_2017}. In order to reach this particular regime, by tuning the individual properties of each nano-oscillator, we 
design an array where the individual frequency of oscillators are regularly 
spaced, and where each oscillator has a synchronization bandwidth to the external input (injection locking range) equal to this 
spacing.  To do this, the frequency $f^{(i)}$ and the injection locking range 
$\Delta_{(i)}$ of all spin-torque nano-oscillators of the array need to be tuned to fulfill the following two conditions:
\begin{equation}
\begin{cases}
 	 (i)  \mid{f^{(i+1)}-f^{(i)}}\mid=\delta_f\pm\epsilon  \\
	 (ii) \: \Delta_{(i)}=\delta_f\pm\epsilon^{'}.
\end{cases}
\end{equation}
$\epsilon$ and $\epsilon^{'}$ are respectively the maximum frequency and 
injection locking range deviations that we tolerate in the choice of our 
individual parameters, here chosen as 5\% of the frequency spacing value 
($\epsilon^{'}=\epsilon=0.05\times \delta_f$). In our approach, $\epsilon$ and $\epsilon^{'}$ are fixed value constraints for which Eq. (5) should be satisfied. Beyond knowing if such constraints can be satisfied or not, those are simply the initial specification of the array we would like to design. From Table II combined with Eq.~(1) and (4) we see that 
the frequency $f^{(i)}$ and the injection locking range $\Delta_{(i)}$ of each 
oscillator $(i)$ can be tuned through three parameters: the free-layer radius, 
its thickness and the applied dc current $\{{\cal R}_{(i)}, L_{(i)}, I_{dc}^{(i)}\}$. We chose 
to separate the individual frequencies with a frequency step $\delta_f$ of 5 
MHz, which corresponds to the typical measured locking ranges for this type of 
oscillators  \cite{romera_vowel_2017}. In order to take into account the 
reachable size accuracy of the nano-dot manufacturing processes, we also impose a minimum dot radius variation between nano-oscillators of $\delta 
{\cal R}=2.0$ nm and a minimum free layer thickness variation of $\delta L=0.$1 nm 
from one nano-dot to another  $\mid{{\cal R}_{(i)}-{\cal R}_{(j)}} |>\delta {\cal R} , 
\mid{L_{(i)}-L_{(j)}} |>\delta L$. These minimum free-layer thickness and radius variations can also be seen as statistical mean indicators that capture the size deviation from nominal realization of the oscillators in the array. In our case, these parameters are given and are part of the initial design problem. Their value will depend on the presence of defects or fabrication imperfections. In this sense they are independent of the specific frequency sensitivity we want to achieve in the array. In order to reach self-sustained oscillations, for given lateral area ($A=\pi {\cal R}^2$) a sufficient electrical current density $J_c=I/A$ needs to be applied. Assuming a constant resistance-area product $\rho$, the static Joule heating in the device will scale linearly with the lateral area: $P_{Joule}=RI^2=(\dfrac{\rho}{A})(J_cA)^2=\rho J_c^2 A$.
Therefore, to avoid large Joule heating due to this area contribution, we consider a maximum nano-dot radius size of 300 nm. Furthermore, 
the maximum and minimum nano-pillar radius (300 and 150 nm) and thickness (8.1 
and 3.0 nm) are chosen in such a way that the magnetic ground state of the FeB 
layer is always a vortex state.

\begin{figure}[ht]
	\includegraphics[scale=0.25]{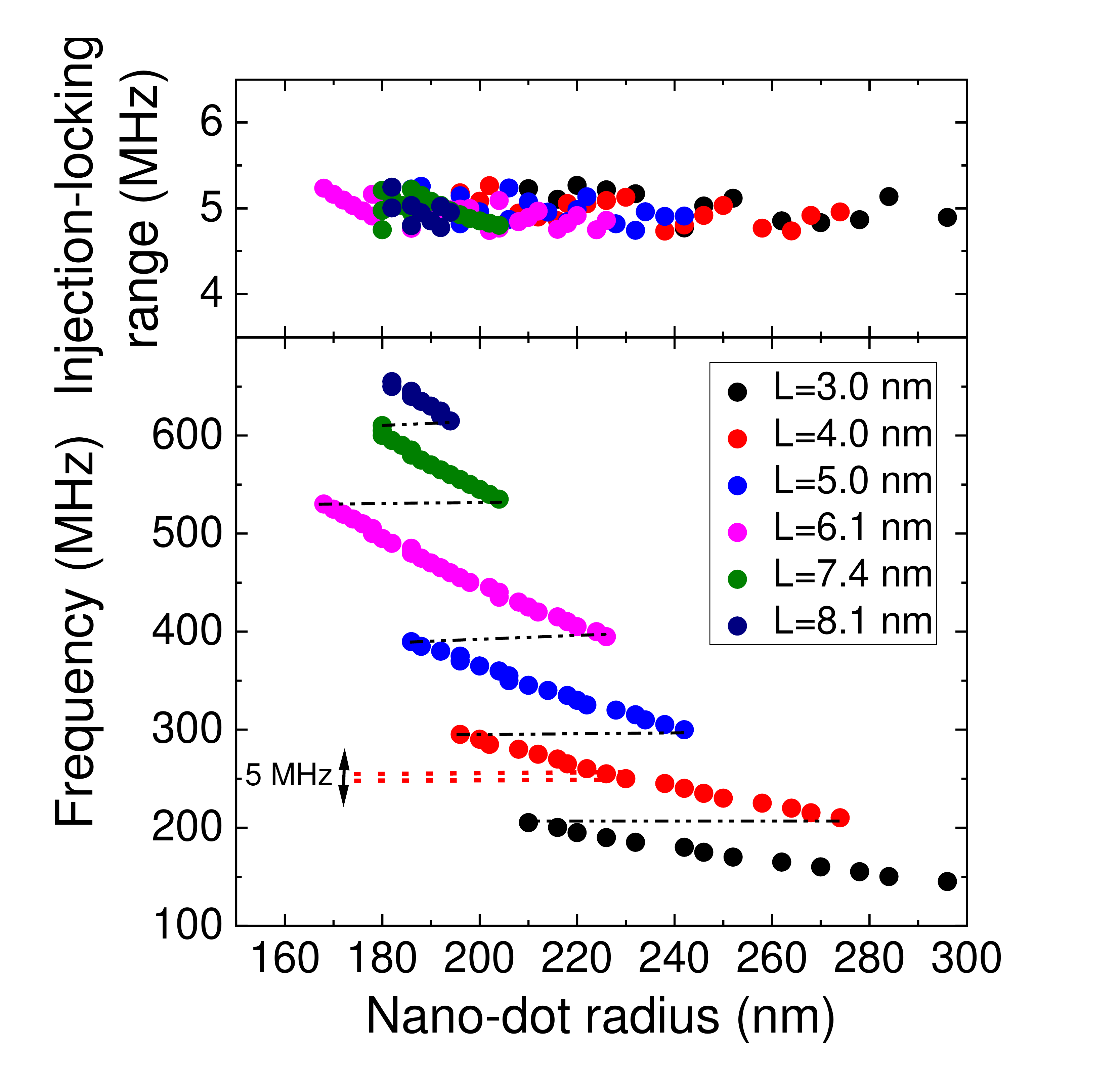}
	\caption{\label{fig:epsart2} Lower graph: analytical auto-oscillation frequency of each nano-oscillator resulting from the application of the selected individual dc current as a function of the chosen nano-dot radius. The color code indicates the corresponding free layer thickness. Upper graph:  distribution of the analytical injection locking range for a constant external microwave signal amplitude, as a function of the chosen nano-dot radius for different thicknesses. The analytical injection locking range remains contained around 5 MHz.
	}
\end{figure}
\begin{figure}[ht]
	\includegraphics[scale=0.25]{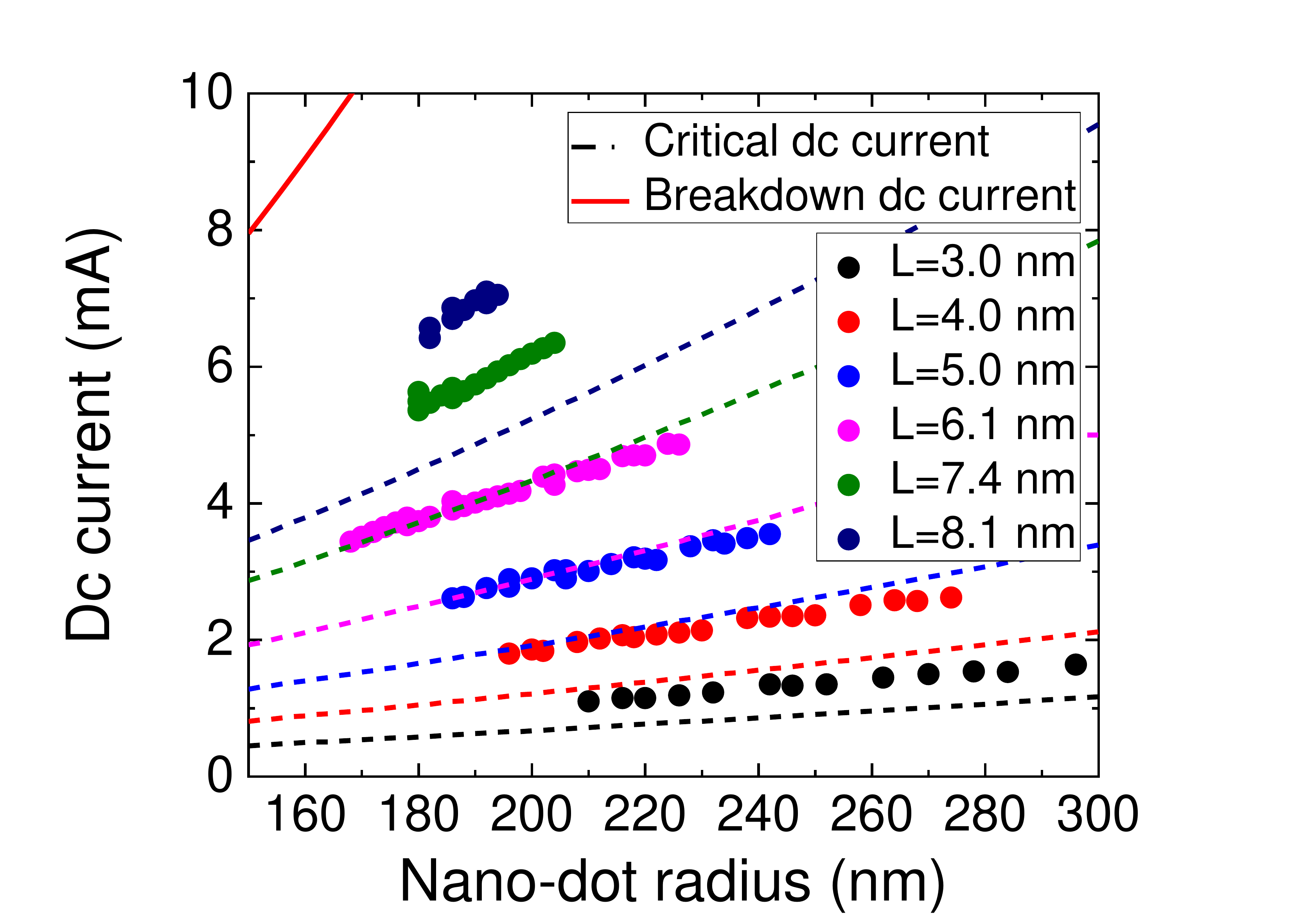}
	\caption{\label{fig:epsart3} Chosen applied dc current versus nano-dot radius for the different selected free-layer thicknesses, dashed lines correspond to the critical current to obtain auto-oscillations. Colors correspond to different free layer thicknesses as in Fig. 2. The red line represents the dc current corresponding to the breakdown voltage. 
	}
\end{figure}

The applied dc currents $I_{dc}^{(i)}$ 
are chosen according to the accuracy of the electrical circuit supplying them. 
Therefore, we impose a minimum current variation of $\delta I=0.1$ mA 
from one oscillator to another one: $\mid{I_{dc}^{(i)}-I_{dc}^{(j)}}\mid>\delta 
I$. The applied dc current $I_{dc}^{(i)}$ (as radius ${\cal R}$ and thickness $L$) is considered as a parameter that allows tuning the individual frequency and injection locking range of the spin-torque oscillator. Its value is maintained constant in time during the microwave information processing. While the constant dc current is applied, all the oscillators of the array are in a self-sustained oscillatory regime. During this stage, the oscillator array receives a collection of magnetic microwave inputs (these inputs can be also encoded as electrical microwave inputs). Depending on the frequency mismatch between the oscillator and the input signal, some spin-torque nano-oscillators in the array can synchronize their oscillations to the input. Figs. 2 and 3 show the calculated values of free-layer radius, 
thickness and applied dc current that fill these constraints as well as 
conditions (i) and (ii) for an array of 100 oscillators. In order to find these values, we explore the three-dimensional space composed by $\{{\cal R}, L, I_{dc}\}$. We defined a grid of points belonging to this finite size space where each point is separated from another by $\delta R$, $\delta L$ and $\delta I$. At each point of this grid, we compute the frequency and the injection locking range (using Eq.~(1,4)). If the computed frequency and injection locking range satisfy the aforementioned conditions of Eq. (5), then the corresponding parameters at that point $\{{\cal R}_{(i)}, L_{(i)}, I_{dc}^{(i)}\}$ are selected to be the parameters of the $(i)$-th oscillator of the array we are designing. If a set of $\{{\cal R}_{(i)}, L_{(i)}, I_{dc}^{(i)}\}$ value gives rise to a frequency and injection-locking range that do not satisfy Eq. (5), this set is excluded and not used to build our array of oscillators. Therefore, for all the arrays presented in this work, all the oscillators verify the condition set by Eq. (5). This restriction can result in a reduction of the number of oscillators that we can have in the array. 

In all panels of Figs. 2 and 3, each dot corresponds 
to one of the 100 oscillators of the array. The bottom panel of Fig. 2 shows 
the auto-oscillation frequencies of the oscillators as a function of their 
radius ${\cal R}_{(i)}$. The corresponding thicknesses $L_{(i)}$  are represented in 
different colors. The resulting frequencies cover a microwave range of 510 MHz 
starting from 145 MHz and ending at 655 MHz. For the considered range of current, the nonlinear frequency shift $\nu^{(i)}$ defined in  Eq.~(3) is comprised  between 9 and 11 which is consistent with values found in the literature \cite{grimaldi2014response}. For vortex-based spin-torque oscillator, the nonlinear frequency shift typically decreases as a function of the applied dc current in the self-sustained regime. If it is too small (respectively too large), the injection locking range become smaller (respectively bigger) too, and therefore the condition (ii) of Eq. (5) is not satisfied. Therefore, the nonlinear frequency shift needs to have an intermediate value. $\nu^{(i)}$ values used here correspond to an intermediate value regime that leads to injection locking ranges around 5 MHz. The top panel of Fig. 2 shows the 
corresponding injection locking ranges of each nano-oscillator. The 
distribution of this injection locking range is narrow around 5 MHz, which means, as desired, that each nano-oscillator of the array has a similar sensitivity 
to the external inputs that it receives. In Fig. 3, the dc currents applied to 
each individual oscillator are shown. Those applied dc currents 
are higher than the critical dc current  $I_{c}^{(i)}$ (dashed 
lines) required to obtain auto-oscillations. This highlights the fact that all the nano-oscillators are in an auto-oscillation regime. In addition, the applied dc 
current is always set smaller than the breakdown current (red straight line) which should not be reached otherwise the magnetic junction would be damaged. Indeed, for bias voltages close or higher than 450 mV, a sudden degradation of tunneling properties is often observed experimentally. Thus, for a given resistance-area product, one can derive the corresponding breakdown current versus nano-pillar radius evolution we used here (see Appendix \ref{app:BreakDownCurr}).

\begin{figure}[ht]
	\includegraphics[scale=0.25]{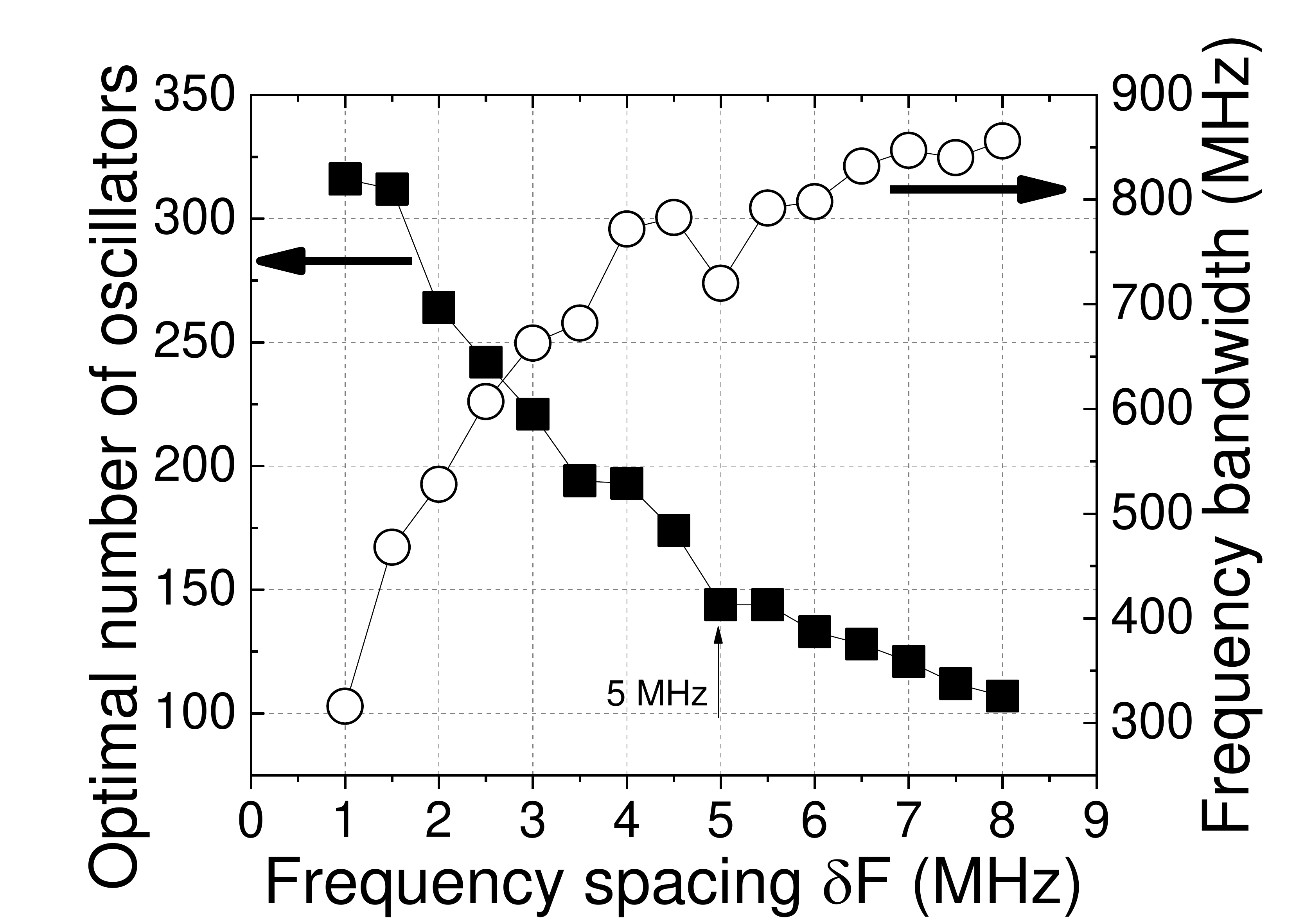}
	\caption{\label{fig:wide4} Maximum number of nano-oscillators (black squares) and their total frequency bandwidth (white circles) in the array as a function of the frequency gap between their auto-oscillation frequencies. For small frequency gaps $\delta_f=1.5$ MHz, arrays of more than 300 nano-oscillators with suitable frequency and synchronization features can be designed.}
\end{figure}

\begin{figure}[ht]
	\includegraphics[scale=0.25]{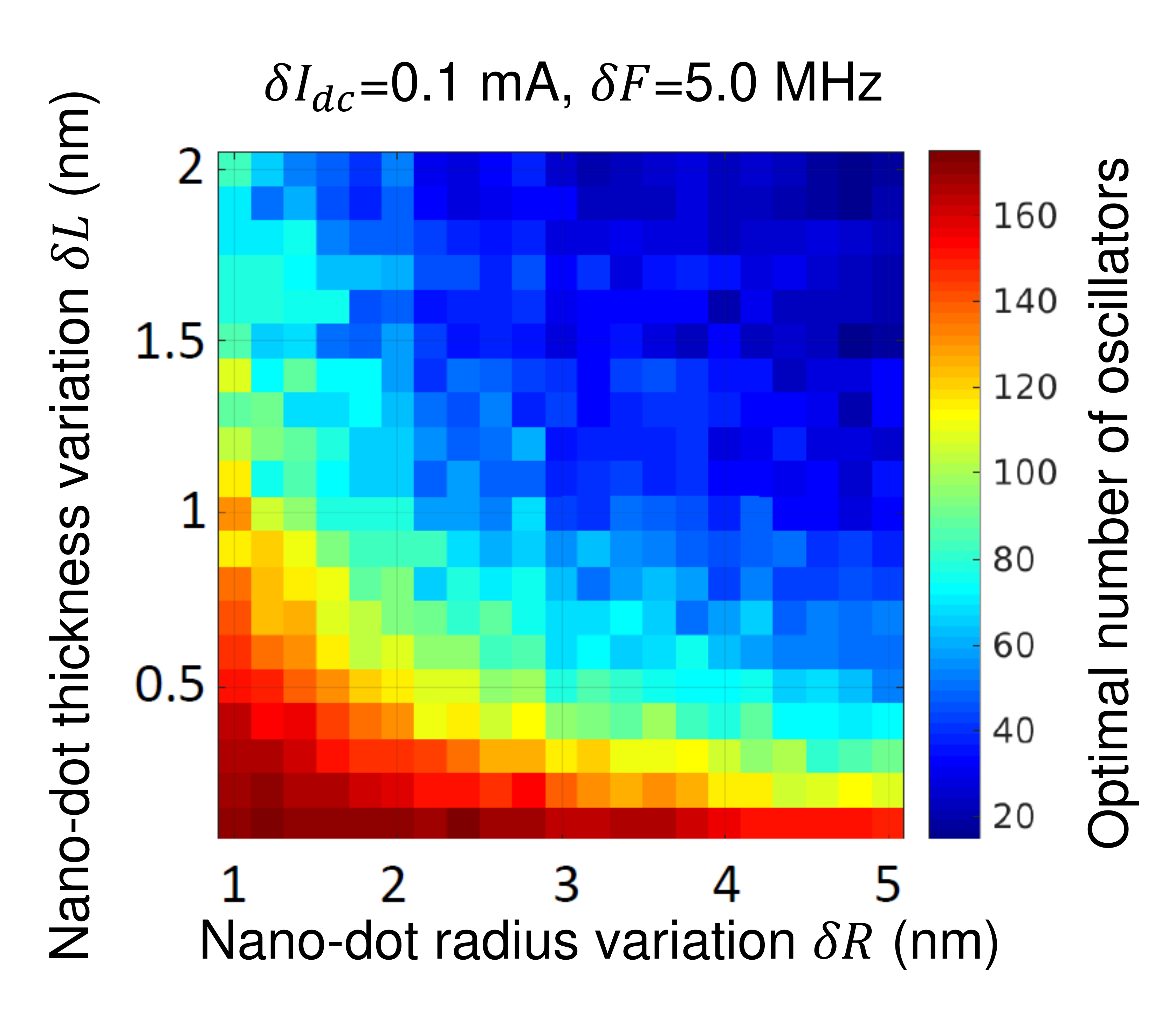}
	\caption{\label{fig:wide5} Colormap showing the maximum number of nano-oscillators in the array as a function of the minimum radius and thickness variations allowed for the nano-pillars, for a minimum dc current variation of $\delta I=0.1$ mA.}
\end{figure}

We now explore the conditions to obtain 
larger arrays ($N>100$) using Eq.~(1,2,4), while insuring frequency and synchronization 
requirements (i) and (ii). Importantly, those conditions were examined for constraints given by the minimum variations of the free-layer size ($\delta {\cal R},\delta L$) and applied dc current $\delta $I. In Fig. 4, we vary the frequency spacing $\delta_f$ between the individual oscillator frequencies from 1.0 to 8.0 MHz. This frequency spacing sets the sensitivity of the array to microwave inputs. For each value of frequency spacing, we computed the optimal number of oscillators in the array (black squares in Fig. 4) that gives rise to the largest frequency bandwidth over which the array will respond (white circles in Fig. 4). Here by array response we mean that at least one oscillator of the array will leave its free-running oscillation frequency and synchronize its oscillation to the frequency of the input frequency that was sent to the array. For a higher or lower input frequency, a different oscillator of the array should react. As for the design of the 100-oscillator array, here we explore the finite $({\cal R}, L, I_{dc})$ parameter space using a three-dimensional grid where each point is regularly separated by $\delta {\cal R}$, $\delta L$ and $\delta I$. Among these grid points, a proportion of them will satisfy frequency and synchronization bandwidth conditions (i) and (ii).  These points are selected to be the parameters of the individual oscillators of the array we design. For a given frequency spacing, the total number of these distinct oscillators corresponds to the optimal number of oscillators than can be found. The observed optimal number of oscillator trends in Fig. 4 can be understood as follows. When the frequency spacing is large, multiple oscillators will satisfy these conditions despite the constraints, here set to $\delta {\cal R}=2.0$ nm,  $\delta L=0.1$ nm and $\delta I=0.1$ mA. Therefore, for the larger frequency spacing $\delta_f$ of 8 MHz, the frequency bandwidth of the array (810 MHz) is practically equal to the frequency range accessible to vortex oscillators (850 MHz), limited only by the vortex ground state stability and the maximum current that can be sustained. The optimal number of oscillators in the array, 110, is then close to the frequency range achievable by vortex oscillators (850 MHz), divided by the frequency gap between oscillators (8 MHz). The situation is different when the frequency spacing between oscillators decreases. Then, due to the precision of lithography and injected current, it becomes more and more difficult to find oscillators with the required frequencies. For a frequency spacing of 1 MHz, the optimal number of oscillators in the array, 330, is much lower than the value that could be achieved without constraints (~close to 800). For this reason, as the array sensitivity increases (smaller $\delta_f$), the frequency bandwidth of the array decreases down to 300 MHz. The number of oscillators shown in Fig. 4 is the optimal array size. If the number of oscillators is smaller, the overall array bandwidth decreases. If the number of oscillators is higher, the required sensitivity is not achieved. To have a larger frequency bandwidth for a given $\delta_f$, smaller minimum variations compared to the one chosen here are required. Fig. 5 shows the calculated optimum number of nano-oscillators in the array with 
the following dc current and frequency constraints: $\delta I=0.1$ mA and 
$\delta_f=5.0$ MHz. The red region corresponding to arrays larger than one 
hundred nano-oscillators are obtained for conditions where the allowed minimal 
variation on radius and thickness are the smallest ones ($\delta {\cal R}<2$ nm and 
$\delta L<0.5$ nm). The blue region corresponding to arrays smaller than forty nano-oscillators can be achieved for less severe constraints where minimal variation on radius and thickness are allowed to be larger ($\delta {\cal R}>2$ nm and $\delta L>1$ nm).

\section{Numerical study of a large spin-torque nano-oscillator array in presence of mutual electrical interaction}

This section considers the impact of mutual interaction between oscillators. The 
largest arrays are obtained for the smallest frequency spacing, for which 
spin-torque nano-oscillators are more coupled if they interact. Electrical connections between oscillators for instance can lead to high mutual interactions when the frequency difference becomes small, of the order of 1 MHz for the oscillators modeled here \cite{tsunegi,Lebrun_Mutual,romera_vowel_2017}. This modifies their oscillation frequency and can lead to their mutual 
synchronization which can affect their ability to be synchronized to an external 
microwave input \cite{romera_enhancing_2016}. 
Here, the coupling phenomena is due to the sum of the rf emissions of the individual oscillators. In our case, where we consider N oscillators in series (see Fig. S1), every individual emission is shared through the whole electrical line. Therefore, every oscillator sees the same sum of total individual electrical microwave emissions. In this sense, all the rf emissions guided by serial electrical connections give rise to a global all-to-all coupling \cite{georges_impact_2008}. This collective coupling effect is not captured by the 
analytical description that we considered until now. We 
now examine the impact of oscillator mutual couplings on the array behavior 
through numerical simulations. For this purpose, we first study the collective 
behavior of an array of 100 electrically coupled spin-torque nano-oscillators that receives the sum of two distinct external microwave magnetic fields. This approach for applying microwave inputs to spin-torque nano-oscillators was used in experiments and simulations for an array of four coupled spin-torque nano-oscillators \cite{romera_vowel_2017}. The
parameters of all nano-oscillators in the array are the ones determined and 
displayed in Fig. 2 and 3. The electrical coupling between nano-oscillators 
resulting from their microwave emissions is described as an additional common 
alternating current that goes through all 
nano-oscillators \cite{georges_impact_2008} 
\begin{equation}
I_{rf}^{com} =\dfrac{1}{Z_0+\sum_{i=1}^{N}R_i}\sum_{i=1}^{N}\lambda \Delta R_i I_{dc}^{i}y_i.
\end{equation}
 Here  $\Delta R_i$ is the mean resistance variation due to the vortex core 
 gyrotropic motion and tunnel magnetoresistance, $Z_0$ is the load 
 impedance which is equal to 50 $\Omega$, $R_i$ is the resistance of the 
 junctions and $\lambda=2/3$ \cite{guslienko2002eigenfrequencies}. In order to 
 obtain the magnetization dynamics of the nano-oscillators, 
 we numerically implement a fourth order Runge-Kutta scheme and solve the 
 coupled differential Thiele equations (7)
\begin{equation}
{\bf{G_i}}\times\dfrac{d{\bf{X_i}}}{dt}-{\bf{D_i}}({\bf{X_i}})\dfrac{d{\bf{X_i}}}{dt}-\dfrac{\partial W_i(I_{rf}^{com})}{\partial {\bf{X_i}}}+{\bf{F_i^{STT}}}(I_{rf}^{com})=0
\end{equation}
simultaneously for the 100 vortex $i=1,2,..100$. Here, ${\bf{X_i}}=(x_i, y_i)$ 
is the vortex core position, ${\bf{G_i}}$ is the gyrovector, ${\bf{D_i}}$ is the 
damping, $W_i$  is the potential energy of the vortex, $\bf{F_i^{STT}}$ is the 
spin-transfer force. The same numerical framework including the mutual 
electrical coupling has been shown previously to reproduce quantitatively the 
synchronization state features observed experimentally for an array of two 
\cite{romera_enhancing_2016} and four \cite{romera_vowel_2017} coupled 
nano-oscillators in presence of external microwave stimuli.

 Fig. 6a shows the large variety of synchronization states obtained when two 
 distinct external microwave stimuli with frequencies $(f_A,f_B)$ are injected 
 to the array of one hundred spin-torque nano-oscillators. By sweeping the 
 frequency of these external stimuli in the frequency range covered by the 
 nano-oscillator array from 145 MHz to 655 MHz, each nano-oscillator is 
in turn synchronized around its 
 free-running auto-oscillation frequency then desynchronized from the external signal. Each square corresponds to 
 one unique synchronization state. The color of squares are chosen arbitrarily to help to distinguish between synchronization states neighbors. In this configuration, 9900 different 
 synchronization states can be reached (by comparison, previous experimental  
 work with four coupled nano-oscillators showed only 20 synchronization 
 states \cite{romera_vowel_2017}). As shown in the synchronization map of Fig. 
 6a and corresponding zoom in Fig. 6b, the individual
 injection locking ranges and the frequency gap between closest nano-oscillator 
 frequencies are very similar and, as designed, have a frequency size deviation 
 smaller than 5\%. This deviation from the desired frequency features ((i) and 
 (ii)) varies with the collective electrical coupling conditions. 
 
 \begin{figure}[ht]
	\includegraphics[scale=0.22]{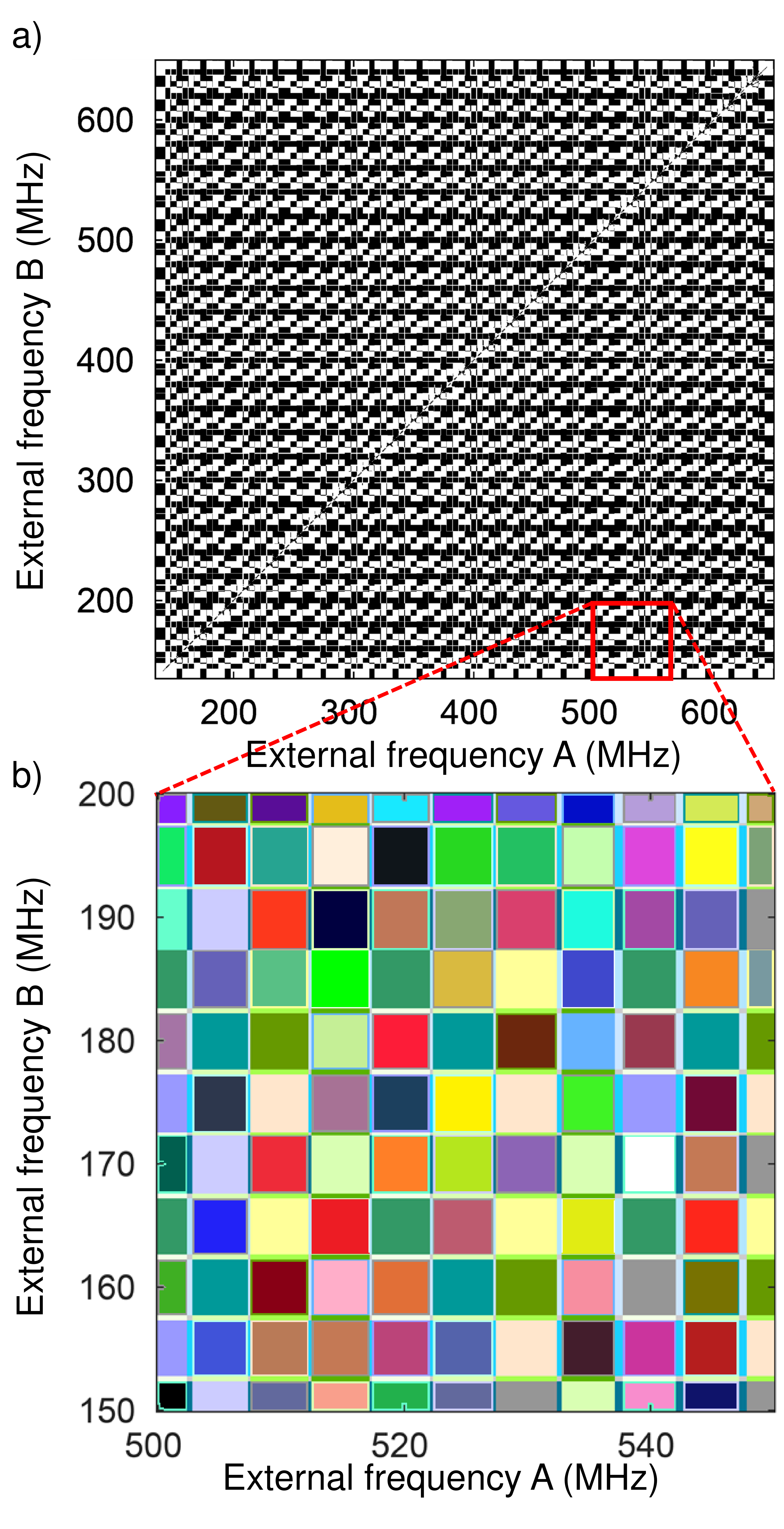}
	\caption{\label{fig:epsart4} a) Simulated synchronization state map of the 100 coupled nano-oscillators with analytically designed free-layer dimensions and applied dc current. The x and y axis correspond to the frequencies of the two microwave inputs injected electrically to the array. Depending on these input frequencies, different oscillators are phase-locked to one of the two microwave inputs. Each small square represents one particular synchronization state.  b) Zoom on a square area of the main synchronization map. Colors are chosen arbitrarily to help to distinguish between synchronization states neighbors.}
\end{figure}
 \begin{figure}[ht]
	\includegraphics[scale=0.25]{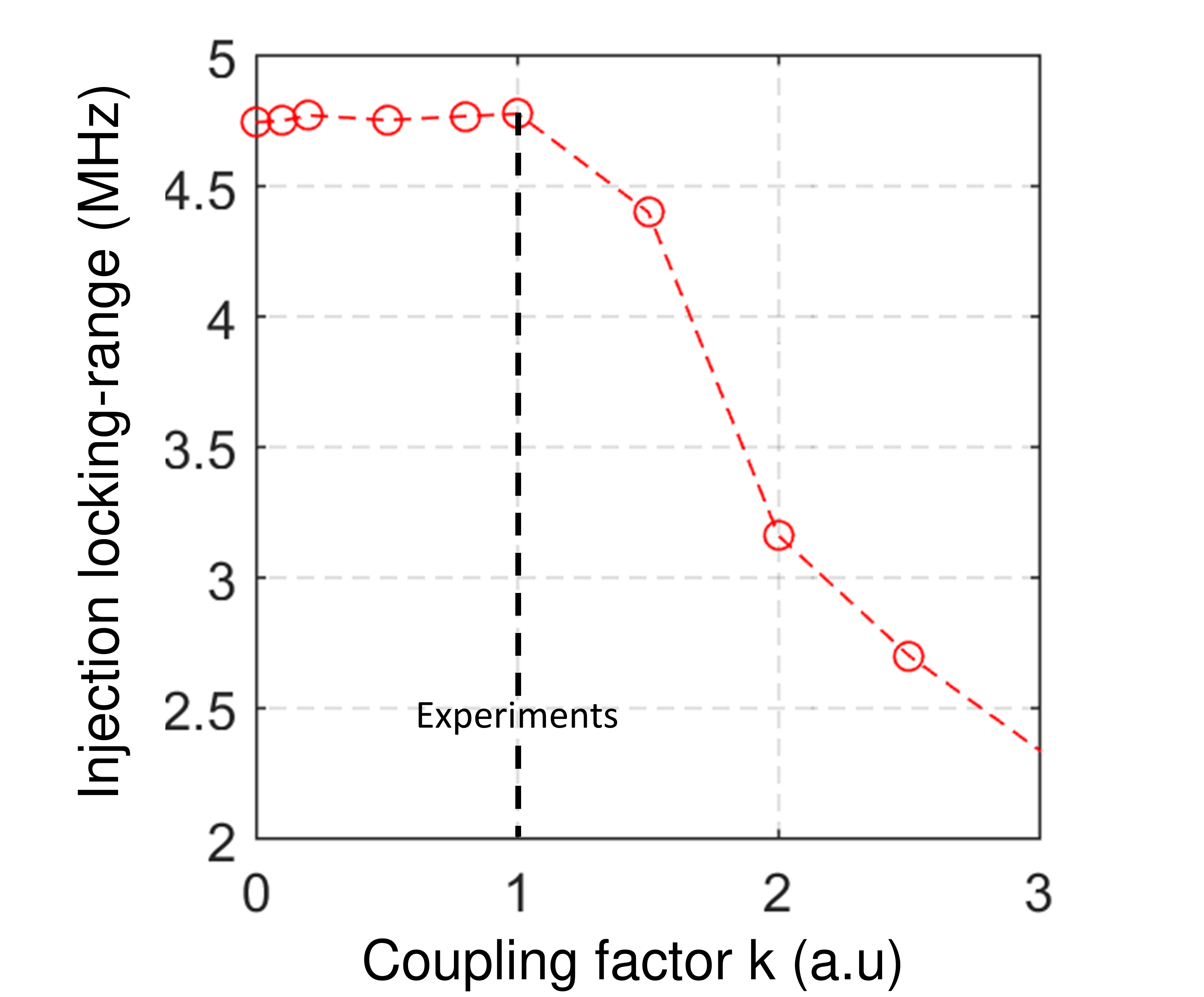}
	\caption{\label{fig:epsart5} Mean injection locking range of 10  simulated coupled spin-torque nano-oscillators as a function of the coupling factor amplitude $k$.}
\end{figure}
 To highlight 
 this effect, we study a smaller array of 10 coupled spin-torque 
 nano-oscillators in presence of two injected microwave signals and simulate 
 the system as it was done for the array of 100 nano-oscillators, for different amplitudes of the electrical coupling. To simulate distinct electrical coupling 
 environments, we multiply the common emitted microwave current generated by 
 all spin-torque nano-oscillators $I_{rf}^{com}$ by a factor $k$ that allows tuning the strength of coupling in simulations. Thus we consider the following new common microwave current  
 $I_{rf}^{com}=k.I_{rf}^{com}$.  $k=1$ corresponds to standard experimental conditions, while $k=0$ corresponds to an uncoupled oscillator array. For the coupling tuning values $k$ used here, different initial positions of the vortex core give rise to the same final frequency in the steady state oscillatory regime. As shown in Fig. 7, when 
 the mutual coupling between oscillators increases, it strongly modifies the ability of the oscillators to phase lock to external microwave signals. For high coupling regimes corresponding to $k>1.5$, the synchronization response of the array drastically decreases 
 from the initially designed one. For such strong electrical coupling conditions, the 
 spin-torque nano-oscillator array will not be sensitive to all the range of input frequencies it was designed for. For standard experimental coupling conditions 
 corresponding to $k=1$, the observed injection locking range  does not decrease, and the array functions remain as desired. This numerical result shows that the analytical 
 approach to designing large size arrays that we propose is robust to  electrical coupling effects.
   
\section{Conclusion}
Our analytical study shows that the properties of the free layer of spin-torque nano-oscillators as well as the amplitude of the injected dc current can be tuned to design arrays of oscillators responding to wide input frequency bandwidths of several hundreds of MHz. The technological constraints on junction dimensions and injected current, due to 
nano-processing and electrical circuit design, impose the optimal size of 
spin-torque nano-oscillator arrays for a given frequency sensitivity. We have shown that the maximum size of an optimal array is around 300 spin-torque nano-oscillators for realistic manufacturing 
parameters and vortex free-layer configuration. Finally, we have shown numerically that the mutual coupling between oscillators does not decrease the array performance as long as coupling remains moderate, close to the experimental values measured for electrical couplings. 

Increasing the overall frequency response from hundreds of MHz to several GHz can be achieved by working with higher frequency junctions than vortex oscillators, for example using oscillators based on uniform magnetization dynamics \cite{bonetti_spin_2009}. The design rules and methods developed here can be easily extended to other types of spin-torque nano-oscillators. The equations for the dynamics of vortex oscillators are indeed identical to the formalism describing spin-torque nano-oscillators in general \cite{slavin_nonlinear_2009,grimaldi2014response}. Beyond magnetic tunnel junction structures, our design approach can be adapted to a novel variety of spintronic devices emerging from the recent spin-orbit torque advances \cite{cai2017electric, li2019manipulation}. Moreover our work can also straightforwardly be extended to uneven frequency spacing between oscillators, following for example a logarithmic scale.

 In summary, we have shown through simulations the possibility to build a 
 device made of a large array of electrically coupled spin-torque 
 nano-oscillators able to respond to microwave signals with a wide range of input frequencies 
 with a constant sensitivity in the whole operating bandwidth. These results 
 open the path to using such arrays in  applications such as spectral analysis, microwave sensing and brain-inspired computing.





\section*{Acknowledgements}
This work was supported by the European Research Council ERC under Grant bioSPINspired 682955. F.A.A. is a Research Fellow of the F.R.S.-FNRS.

\nocite{*}

\appendix


\section{Oersted field and magnetostatic field contributions}
\label{app:OerstedMagnetoStat}

The Oersted field ${\bf{B_{Oe}}}$ (due to the perpendicular flow of electrical charge current J) modifies the energy landscape seen by the magnetic vortex core in the free-layer. For a given vortex core position ${\bf{X}}=(x,y)$, this additional energy contribution $E_{Oe}$ is captured by a Zeeman energy term, formed by the interaction between the local Oersted field ${\bf{B_{Oe}}}({\bf{r}})$ and the local magnetization ${\bf{m}}({\bf{r}},{\bf{X}})$, averaged over the whole magnetic volume of the free-layer. Here  ${\bf{r}}$  corresponds to the position of the local magnetization. For a given vortex magnetization profile, this energy term was calculated and approximated by a polynomial expression \cite{S_RefA,khvalkovskiy_nonuniformity_2010}.
\begin{multline}
   E_{Oe}(X)=-\int{ {\bf{B_{Oe}}}({\bf{r}}){\bf{m}}({\bf{r}},{\bf{X}}) d{\bf{r}}  }\\
   = J (\dfrac{1}{2} \kappa_{Oe} X^2 +\dfrac{1}{4} \kappa_{Oe}^{'} \dfrac{X^4}{R^2}) + O(\dfrac{X^6}{R^4}).
\end{multline}
In this expression, the linear Oersted field confinement coefficient $\kappa_{Oe}$ corresponds simply to a scaling factor of the term in $X^2$ (respectively the nonlinear Oersted field confinement coefficient $\kappa_{Oe}^{'}$ scales the term in $X^4$). Here $R$ is the nano-pillar radius of the free-layer.

Similarly, for a given vortex core position ${\bf{X}}$, one can derive the Zeeman interaction energy term due to the demagnetization field ${\bf{B_{demag}}}$. This demagnetization field is due to spatial magnetic charges caused by the intrinsic magnetization distribution in the free-layer than can appear in the absence of an applied bias. For a given vortex magnetization profile, Gaidedei et al. \cite{gaididei_magnetic_2010} calculated this energy term which is approximated through the following polynomial expression:

\begin{multline}
   E_{ms}(X)=-\int{ {\bf{B_{demag}}}({\bf{r}}){\bf{m}}({\bf{r}},{\bf{X}}) d{\bf{r}}  }\\
   = \dfrac{1}{2} \kappa_{ms} X^2 +\dfrac{1}{4} \kappa_{ms}^{'} \dfrac{X^4}{R^2} + O(\dfrac{X^6}{R^4})
\end{multline}

In this expression, the linear magnetostatic confinement coefficient $\kappa_{ms}$   corresponds to a scaling factor of the term in $X^2$ (respectively the nonlinear magnetostatic confinement coefficient $\kappa_{ms}^{'}$ scales the term in $X^4$).
\\
\section{Breakdown current}
\label{app:BreakDownCurr}

For a given resistance-area product $\rho$, the following expression defines the breakdown current as a function of the nano-pillar radius ${\cal R}$:
\begin{equation}
I_{breakdown}=\dfrac{\pi {\cal R}^2 V_{breakdown}}{\rho}
\end{equation}

Here, $V_{breakdown}=450$ mV, $\rho=4.0\times10^{-12}$ $\Omega m^2$ is chosen to be similar to the FeB based (free-layer) spin-torque oscillators used experimentally in our experimental work of \cite{romera_vowel_2017}. The evolution of this breakdown current is plotted in Fig. 3.

\section{Electrical circuit of the studied spin-torque nano-oscillator array}
\label{app:ElectricalCircuit}

 \begin{figure}[ht]
	\includegraphics[scale=0.26]{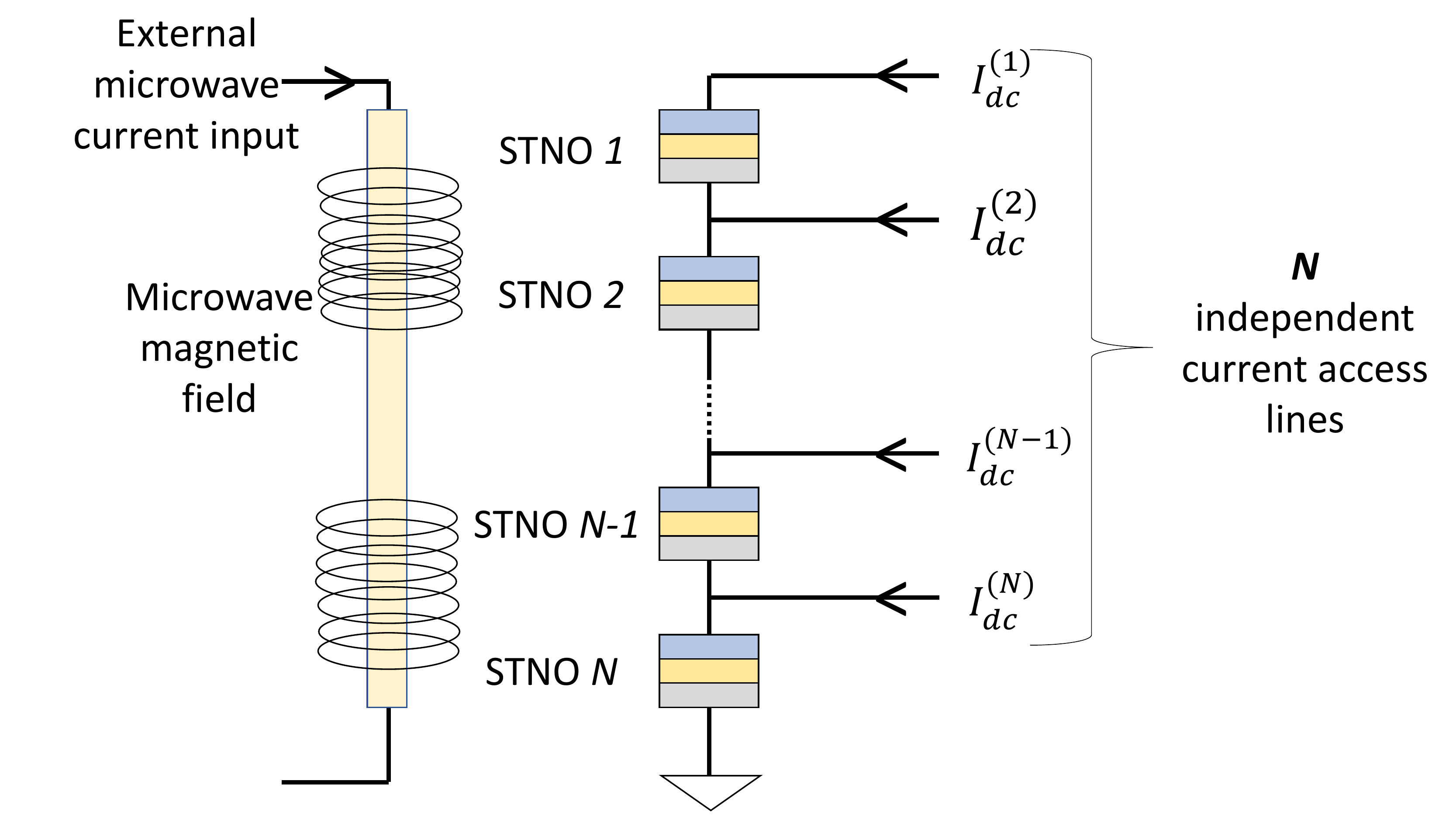}
	\caption{\label{fig:epsart4S} Schematic of the electrical circuit of an array of N spin-torque nano-oscillators receiving N independent dc currents. The array receives the microwave input in the form of an ac magnetic field generated by an inductive line.}
\end{figure}


\providecommand{\noopsort}[1]{}\providecommand{\singleletter}[1]{#1}%
%



\end{document}